\pdfoutput=1
\documentclass[11pt]{article}

\usepackage[utf8]{inputenc}
\usepackage[T1]{fontenc}
\usepackage{lmodern}
\usepackage{float}
\usepackage{amsmath, amssymb}
\usepackage{graphicx}
\usepackage{tikz}
\usetikzlibrary{arrows.meta, positioning, calc, backgrounds, fit, shapes.geometric}
\usepackage{standalone}
\usepackage{xcolor}
\definecolor{tealDark}{HTML}{043A71}
\definecolor{tealMid}{HTML}{0D4E8B}
\definecolor{tealLight}{HTML}{B8D4E8}
\definecolor{tealPale}{HTML}{E8F0F8}
\definecolor{slateDark}{HTML}{021D3A}
\definecolor{slateMid}{HTML}{0A2D52}
\definecolor{sageDark}{HTML}{1A6FB5}
\definecolor{sageLight}{HTML}{6BAED6}
\definecolor{grayLabel}{HTML}{4A6278}
\definecolor{grayDim}{HTML}{7A8D9C}
\definecolor{grayArrow}{HTML}{5B7088}
\definecolor{bandBorder}{HTML}{A3C4DC}

\usepackage[a4paper,margin=1in]{geometry}
\usepackage[expansion=false]{microtype}
\usepackage[hidelinks]{hyperref}
\usepackage{orcidlink}
\newcommand{\hiddennotice}[1]{}
\usepackage{svg}
\svgpath{{Figures/}}

\newcommand{\Q}{\mathbf{Q}}
\newcommand{\dPDF}{3D-\ensuremath{\Delta}PDF}

\title{Connecting Diffuse Scattering to Atomic-Site-Resolved Occupancy and Displacement Fields through Fourier Filtering}

\author{%
    \textbf{Maksim Eremenko}\,\orcidlink{0000-0002-2875-968X}$^{1,2,3,*}$, \textbf{Victor Krayzman}\,\orcidlink{0000-0001-9010-2681}$^{1}$,\\
    \textbf{Matthew G. Tucker}\,\orcidlink{0000-0002-2891-7086}$^{2}$, \textbf{Igor Levin}\,\orcidlink{0000-0002-7218-3526}$^{1}$\\[0.5em]
    \small $^{1}$Materials Measurement Science Division, National Institute of Standards and Technology,\\
    \small Gaithersburg, MD, USA\\
    \small $^{2}$Spallation Neutron Source, Oak Ridge National Laboratory, Oak Ridge, TN, USA\\
    \small $^{3}$Theiss Research, La Jolla, CA, USA\\[0.5em]
    \small $^{*}$Corresponding author: eremenkom@ornl.gov
}

\date{} 

\begin{document}
\maketitle

\hiddennotice{Notice: This manuscript has been authored by UT-Battelle, LLC, under contract DE-AC0500OR22725 with the US Department of Energy (DOE). The US government retains and the publisher, by accepting the article for publication, acknowledges that the US government retains a nonexclusive, paid-up, irrevocable, worldwide license to publish or reproduce the published form of this manuscript, or allow others to do so, for US government purposes. DOE will provide public access to these results of federally sponsored research in accordance with the DOE Public Access Plan.}

\begin{abstract}
Local structural correlations are encoded in diffuse scattering, but identifying atomic motifs that produce specific diffuse features can be challenging.  We introduce MOSAIC, a computational framework for this task when an atomistic configuration is available and a phase-bearing scattering amplitude can be calculated. Our approach relies on applying the Fourier filter to this amplitude over the reciprocal-space regions encompassing the scattering features of interest to obtain maps of atomic displacements and site occupancies responsible for those features.    The method is effective in interrogating the nature and spatial distributions of interatomic correlations in large-scale structural models, such as obtained using Reverse Monte Carlo refinements from experimental data, molecular dynamics, or Monte Carlo simulations, or 2D structural projections derived from atomic-resolution electron microscopy images. 
\end{abstract}

\section{Introduction}
Functional responses in crystalline materials -- dielectric, catalytic, sensing, and thermal -- are often governed by local deviations from the average atomic order. These deviations tend to be correlated, and the resulting correlated disorder generates characteristic diffuse-scattering patterns in X-ray, neutron, or electron diffraction.  The shapes and intensity distributions of this diffuse scattering encode information on local displacement fields, chemical ordering, and strain textures that are largely inaccessible to traditional crystallographic analysis using Bragg diffraction~\cite{krivoglaz2012x}.  A significant challenge is not only to establish the types and spatial attributes (extent, dimensionality, topology) of these correlations but also to understand how they evolve across length scales to produce the observed average structure.   Developing methodologies capable of robustly extracting such information from diffuse scattering remains a significant frontier in structural characterization.

Several methods have been used to derive atomistic structural models from diffuse-scattering intensities, including direct Monte Carlo (MC) modeling of these data based on assumed atomic
interactions~\cite{butler1992calculation,welberry1994interpretation} and Reverse
Monte Carlo (RMC) refinements of atomic arrangements by fitting the observed diffuse
features~\cite{mcgreevy2001reverse}.  Alternatively, the Fourier transform (FT) of the diffuse intensity provides a three-dimensional difference pair distribution function (\dPDF) that highlights correlated deviations from the average structure and can also serve as a target for structural refinements~\cite{weber2012three}.  However, even when a structural configuration reproduces diffuse scattering, it often remains challenging to identify which specific correlated motifs within that structure give rise to particular diffuse features and how these motifs are spatially arranged and interconnected.  Similar interpretational difficulties arise for atomistic configurations obtained from theoretical simulations using molecular dynamics (MD) or MC methods: while these models can reproduce diffuse scattering, its structural origins frequently remain unclear.

A closely related situation arises in atomic-resolution transmission electron microscopy (TEM).  Modern aberration-corrected scanning TEM (STEM) and ptychographic phase reconstructions provide direct projections of atomic-column chemical occupancies and displacements with sub-angstrom precision and depth resolution.  A common approach for analyzing such images has been to compute the fast FT (FFT) of the image intensity.  If superlattice peaks or diffuse scattering between Bragg peaks are present, they can be selectively inverse-Fourier transformed to generate filtered images that highlight regions contributing to the selected features.  These images display positive and negative contrast modulations, reflecting deviations from the average structural motif, which effectively visualize ordering patterns and their spatial distributions; however, the analysis remains largely qualitative and does not yet provide a quantitative means to determine the underlying chemical or displacement correlation fields.  

Here, we present a computational framework that uses Fourier filtering of the scattering amplitude calculated for a known atomic configuration to identify the site-occupancy and displacement fields responsible for the selected scattering features.   The workflow starts with an atomic configuration, from which the diffuse scattering amplitude, \(\Delta A(\Q)\), is computed (by subtracting the average-structure term from the total amplitude \(A_{\mathrm{tot}}(\Q)\)). The diffuse features of interest are then isolated by applying phase-preserving masks, and the inverse FT is applied to the masked amplitude, with the transform calculated around each atomic site.  The resulting data are analyzed to link the selected scattering features with the underlying atomic displacements and site-occupancy variations.  This method interrogates a supplied structural model or projection to compute a phase-bearing amplitude; it does not reconstruct a unique atomistic structure directly from intensity-only diffuse-scattering data. Previously, we successfully applied a related feature-filtering analysis to identify chemical-ordering and displacement patterns in atomic configurations derived from RMC refinements of several complex oxides ~\cite{Eremenko2019,eremenko2025emergent,krayzman2022incommensurate}.  Now, we have developed it into a quantitative framework with dedicated software that determines both the relevant patterns and the magnitudes of deviations from the average structure that give rise to specific scattering features.

Our framework is designed to work with large atomic ensembles and calculated diffuse-scattering datasets that span extensive regions of reciprocal space.  We employ the nonuniform fast Fourier transform (NUFFTs) to compute the scattering amplitude on arbitrary \(\mathbf Q\)-point meshes and to efficiently invert selected regions of reciprocal space. This approach enables high throughput while accurately treating experimental geometries with targeted sampling around diffuse features of interest. Recent NUFFT implementations achieve high numerical accuracy and strong scaling across advanced CPU and GPU architectures, making amplitude-space filtering practical for data volumes typical of contemporary diffuse scattering measurements~\cite{greengard2004accelerating,barnett2019parallel,shih2021cufinufft}.

The paper is organized into three main parts. First, we establish a strictly linear phase-preserving pipeline from \(\Delta A(\Q)\) to atom-centered real-space fields, ensuring exact additivity across arbitrary reciprocal-space mask partitions. Second, we derive and validate linear estimators constructed from the even (symmetric) and odd (asymmetric) components of the inverse-transformed amplitude, which recover picometer-scale displacements for each atomic site and isolate chemical contrast. Third, we demonstrate that the same operators apply to model-conditioned analysis of atomic configurations, whether simulated or obtained via RMC refinements against experimental data.  

\section{Methods: Theoretical Framework}
\label{sec:methods}

\begin{figure}[!htbp]
  \centering
  \includegraphics[width=\linewidth]{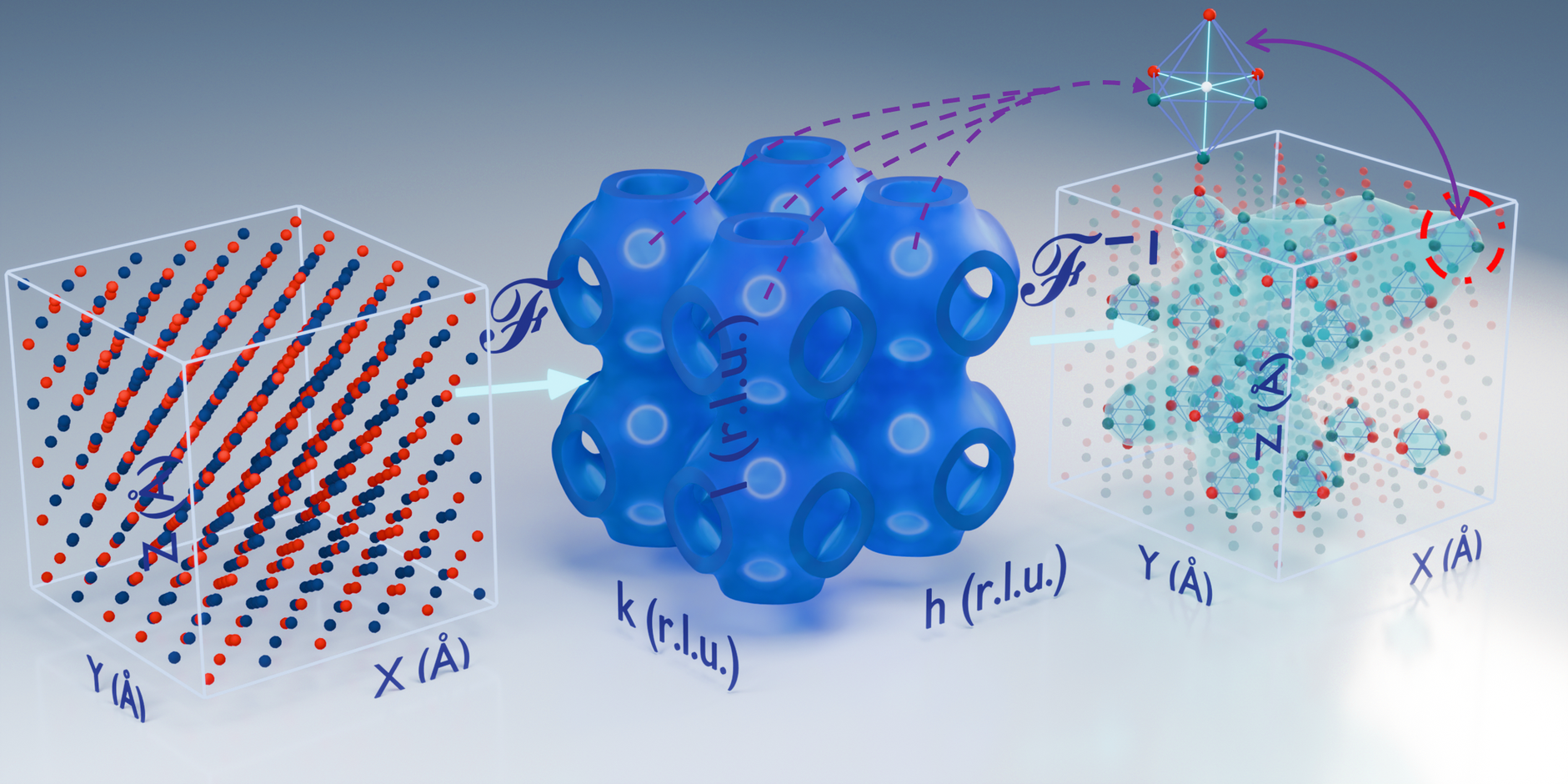}
  \caption{Conceptual illustration of MOSAIC. Starting from an atomistic configuration (left), the complex scattering amplitude is calculated in reciprocal space (center). Features of interest are isolated using masks (light-shaded disks), and the masked amplitudes are Fourier-transformed back to real space to reconstruct the atomic-occupancy and displacement fields (right) corresponding  to the selected feature.}
  \label{fig:Fig1}
\end{figure}

Here, we consider a scenario where an atomic configuration has been determined from experimental data (e.g., via large-box refinements using the RMC method) or from atomistic simulations (e.g., MD or MC). The atomic coordinates can be used to calculate the scattering amplitude. The issue we address is how to connect specific diffuse-scattering features, if present, to the atomic-site-resolved occupancy and displacement fields in the original configuration. The input is therefore a structural model, not intensity data. In the following, we outline our method, which has been implemented in the analysis software called MOSAIC. The general concept is illustrated in Fig.~\ref{fig:Fig1}.

\subsection{Calculations of scattering amplitude}
\label{subsec:kinematic}

We begin by revisiting the formalism for separating contributions to the scattering amplitude from the average crystallographic structure and local deviations associated with atomic disorder.   In the kinematic approximation~\cite{cowley1995diffraction}, which assumes single scattering events, the total complex amplitude is represented as a coherent sum of plane waves originating from atomic positions. We consider a crystalline supercell composed of $N$ unit cells. The instantaneous position of atom $m$ in unit cell $n$ is written as
\begin{equation}
    \mathbf{r}_{mn} = \mathbf{R}_n + \bar{\mathbf{r}}_m + \mathbf{u}_{mn},
\end{equation}
where $\mathbf{R}_n$ is the lattice vector of the cell $n$, $\bar{\mathbf{r}}_m$ is the average (crystallographic) position of the site $m$ within the $n$-th cell, and $\mathbf{u}_{mn}$ is the local deviation from this average position. The total scattering amplitude at the reciprocal-space vector $\mathbf{Q}$ is then
\begin{equation}
    A_{\mathrm{tot}}(\mathbf{Q}) = \sum_{n=1}^{N} \sum_{m} f_{mn}(\mathbf{Q})
    \exp \bigl[ i\mathbf{Q} \cdot \bigl(\mathbf{R}_n + \bar{\mathbf{r}}_m + \mathbf{u}_{mn}\bigr) \bigr],
    \label{eq:Atot}
\end{equation}
where $f_{mn}(\mathbf{Q})$ is the scattering factor (or scattering length for neutrons) of the species occupying the site $(m,n)$.
Throughout, real-space vectors are expressed in Cartesian coordinates (units of \AA). We use the crystallographic $2\pi$ convention for reciprocal space, with $\mathbf{Q}$ expressed in \AA$^{-1}$. When convenient, we express $\mathbf{Q}$ in reciprocal-lattice coordinates $(h,k,l)$ (reciprocal-lattice units, rlu) as $\mathbf{Q} = h\,\mathbf{a}^* + k\,\mathbf{b}^* + l\,\mathbf{c}^*$.

The coherent average amplitude is defined as
\begin{equation}
    \bigl\langle A(\mathbf{Q}) \bigr\rangle =
    \sum_{m} \bar{f}_m(\mathbf{Q})
    \exp\bigl[ i\mathbf{Q} \cdot \bar{\mathbf{r}}_m \bigr],
    \label{eq:Aavg}
\end{equation}
where $\bar{f}_m(\mathbf{Q})$ is the scattering factor of the site $m$ in the average structure. The corresponding interference function representing the FT of the supercell's finite-size shape is
\begin{equation}
    \psi(\mathbf{Q}) = \sum_{n=1}^{N} \exp\bigl(i\mathbf{Q}\cdot\mathbf{R}_n\bigr).
\end{equation}
The product $\langle A(\mathbf{Q}) \rangle \,\psi(\mathbf{Q})$ represents the contribution of the average structure to $A_{\mathrm{tot}}(\mathbf{Q})$.

The diffuse-scattering amplitude can then be defined as
\begin{equation}
    \Delta A(\mathbf{Q}) =
    A_{\mathrm{tot}}(\mathbf{Q}) - \bigl\langle A(\mathbf{Q}) \bigr\rangle \,\psi(\mathbf{Q}),
    \label{eq:Adiff}
\end{equation}
so that, in the limit of an ideal periodic crystal, the second term reproduces the Bragg scattering and $\Delta A(\mathbf{Q})$ captures contributions from the local disorder ~\cite{butler1992calculation}.

Specific scattering features are linked to the underlying real-space fields via the inverse Fourier transform of $\Delta A(\mathbf{Q})$ (or, where useful, $A_{\mathrm{tot}}(\mathbf{Q})$) over discrete subsets of $\mathbf{Q}$. Let $\mathcal{F}$ denote a chosen set of points in reciprocal space -- for example, those forming a diffuse peak, rod, or lobe. The real-space field associated with this feature is defined as
\begin{equation}
    \mathcal{R}_{\mathcal{F}}(\mathbf{r}) =
    \frac{1}{N_{\mathrm{tot}}} \sum_{\mathbf{Q}}
    W_{\mathcal{F}}(\mathbf{Q})\,B(\mathbf{Q})
    \exp\bigl(-i\mathbf{Q}\cdot\mathbf{r} \bigr),
    \label{eq:psi_feature}
\end{equation}
where $B(\mathbf{Q})$ is the selected amplitude field (typically $B=\Delta A$), $W_{\mathcal{F}}(\mathbf{Q})$ is the window selecting the feature region $\mathcal{F}$ (binary or smoothly tapered) and $N_{\mathrm{tot}}$ is the total number of sampled reciprocal-space points in the full transform volume. Because Eq.~\eqref{eq:psi_feature} is linear in $B(\mathbf{Q})$ and uses a feature-independent normalization $N_{\mathrm{tot}}$, contributions from disjoint selections $\mathcal{F}_1,\mathcal{F}_2,\dots$ add exactly, and the associated real-space fields obey
\begin{equation}
    \mathcal{R}_{\mathcal{F}_1 \cup \mathcal{F}_2}(\mathbf{r}) =
    \mathcal{R}_{\mathcal{F}_1}(\mathbf{r}) + \mathcal{R}_{\mathcal{F}_2}(\mathbf{r})
\end{equation}
whenever $W_{\mathcal{F}_1}(\mathbf{Q})\,W_{\mathcal{F}_2}(\mathbf{Q})=0$ for all $\mathbf{Q}$. This strict linearity underpins all subsequent reconstructions in MOSAIC.

More generally, nonuniform reciprocal-space quadrature can be written with fixed sampling weights
$w_{\mathbf Q}$ multiplying each summand; in the present implementation all sampled
$\mathbf Q$ points are assigned a unit weight, $w_{\mathbf Q}=1$, and this factor is therefore
omitted from Eq.~\eqref{eq:psi_feature} and the formulas below.

The reciprocal-space sampling is chosen to preserve Hermitian symmetry, so that the inverse-transformed field is real-valued up to numerical roundoff; the corresponding complex generalization is described in Supplementary Note~1.

\subsection{Filtering Chemical (Occupational) Ordering}
\label{subsec:chemical_mode}

Chemical ordering is often accompanied by static atomic displacements that accommodate dissimilar sizes and electric charges of species involved, resulting in diffuse scattering features that reflect both effects~\cite{welberry2022diffuse}.

Within the MOSAIC framework, the occupational ordering is separated from the concurrent static displacements by constructing a reference configuration in which all atoms occupy their average crystallographic positions, while the chemical occupancy distribution is maintained exactly as in the original configuration. This defines the Chemical Mode. Starting from the full atomistic model, we set
\begin{equation}
    \mathbf{u}_{mn} \rightarrow \mathbf{0}
\end{equation}
for every site $(m,n)$, but retain the actual species at each site. The instantaneous positions entering the calculation are then
\begin{equation}
    \mathbf{r}^{(\mathrm{chem})}_{mn} =
    \mathbf{R}_n + \bar{\mathbf{r}}_m,
\end{equation}
and the total amplitude for this configuration, having all atoms at the average-structure positions, is
\begin{equation}
    A_{\mathrm{tot}}^{(\mathrm{chem})}(\mathbf{Q}) =
    \sum_{n=1}^N \sum_m f_{mn}(\mathbf{Q})\,
    \exp\bigl[i\mathbf{Q}\cdot(\mathbf{R}_n + \bar{\mathbf{r}}_m)\bigr],
    \label{eq:Atot_chem}
\end{equation}
where $f_{mn}(\mathbf{Q})$ denotes the scattering factor of the species occupying site $(m,n)$. The coherent reference subtracted from $A_{\mathrm{tot}}^{(\mathrm{chem})}(\mathbf Q)$ is the same average-structure term $\langle A(\mathbf Q)\rangle\,\psi(\mathbf Q)$ defined in Eqs.~\eqref{eq:Aavg} and \eqref{eq:Adiff}.

The diffuse amplitude associated with chemical ordering is then defined, using the same Butler--Welberry decomposition ~\cite{butler1992calculation}, as:
\begin{equation}
    \Delta A_{\mathrm{chem}}(\mathbf{Q}) =
    A_{\mathrm{tot}}^{(\mathrm{chem})}(\mathbf{Q})
    - \bigl\langle A(\mathbf{Q}) \bigr\rangle \,\psi(\mathbf{Q}).
    \label{eq:Adiff_chem}
\end{equation}
By construction, displacive contributions are suppressed (all $\mathbf{u}_{mn}=0$ in both terms), and $\Delta A_{\mathrm{chem}}(\mathbf{Q})$ represents the diffuse scattering generated solely by deviations of $f_{mn}(\mathbf{Q})$ from their average values $\bar{f}_m(\mathbf{Q})$. Equivalently,
\begin{equation}
    \Delta A_{\mathrm{chem}}(\mathbf Q)=
    \sum_{n,m}
    \bigl[f_{mn}(\mathbf Q)-\bar f_m(\mathbf Q)\bigr]
    \exp\bigl[i\mathbf Q\cdot(\mathbf R_n+\bar{\mathbf r}_m)\bigr].
    \label{eq:Adiff_chem_explicit}
\end{equation}

To examine how a particular reciprocal-space feature reflects chemical correlations, we apply the same selection procedure as for the full diffuse amplitude. For a chosen region $\mathcal{F}$ in $\mathbf{Q}$-space, we multiply $\Delta A_{\mathrm{chem}}(\mathbf{Q})$ by the window $W_{\mathcal{F}}(\mathbf{Q})$ before computing the corresponding real-space field.
\begin{equation}
    \mathcal{R}_{\mathrm{chem},\mathcal{F}}(\mathbf{r}) =
    \frac{1}{N_{\mathrm{tot}}} \sum_{\mathbf{Q}}
    W_{\mathcal{F}}(\mathbf{Q})\,B(\mathbf{Q})
    \exp\bigl(-i\mathbf{Q}\cdot\mathbf{r} \bigr),
    \label{eq:psi_chem}
\end{equation}
This field is dominated by variations in occupancy at average-lattice sites. In practice, the relevant signal at site $(m,n)$ is captured by the value of $\mathcal{R}_{\mathrm{chem},\mathcal{F}}(\mathbf{r})$,
\begin{equation}
    C_{mn}^{(\mathrm{chem})} \equiv
    \mathcal{R}_{\mathrm{chem},\mathcal{F}}\bigl(\mathbf{R}_n + \bar{\mathbf{r}}_m\bigr),
\end{equation}
which provides a single scalar value per atomic site. For example, in a binary system with species A and B, $C_{mn}^{(\mathrm{chem})}$ is, up to a feature-dependent scale factor, proportional to the local form-factor contrast (via $\Delta f = f_{\mathrm{A}} - f_{\mathrm{B}}$), and thus directly tracks the local deviation of the site’s occupancy from its average value.

Thus, the Chemical Mode emphasizes occupancy- and chemistry-related contributions
to a selected diffuse feature, while deliberately suppressing parts of the signal that
depend on instantaneous displacements $\mathbf{u}_{mn}$. This is a separation
within the supplied model and does not imply that chemical and displacive correlations
are physically independent. The resulting three-dimensional maps of $C_{mn}^{(\mathrm{chem})}$ can reveal chemical nanodomains, compositional modulations, or anti-phase boundaries. The same linear definition, Eq.~\eqref{eq:Adiff}, is used consistently for both chemical- and displacement-sensitive reconstructions; the difference lies only in the configuration from which $A_{\mathrm{tot}}$ and $\langle A\rangle$ are computed (i.e., a disordered configuration versus average-position reference) and in how the resulting real-space fields are interpreted at the atomic-site level.

\subsection{Reconstruction of Atomic Displacements }
\label{subsec:disp_mode}

\begin{figure}[!htbp]
  \centering
  \includegraphics[width=1.0\linewidth]{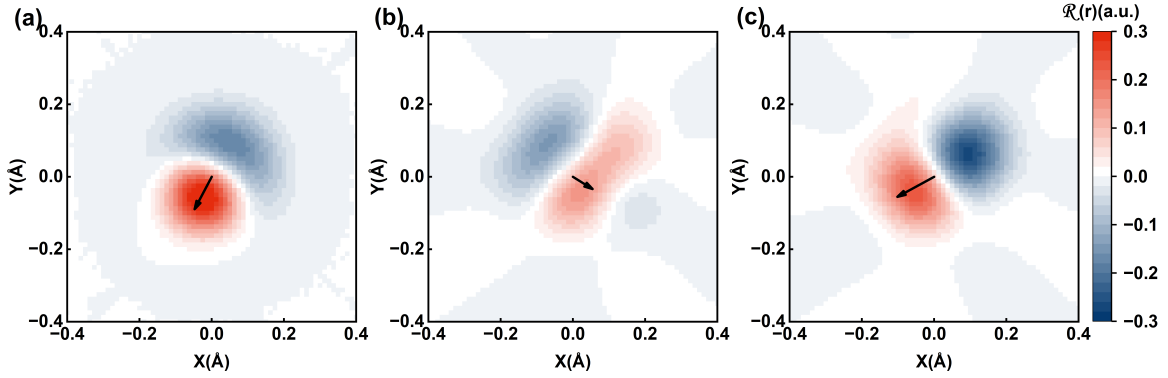}
  \caption{Example of a real-space 2D representation of a local atomic displacement obtained by reciprocal-space filtering. Each panel shows \(\mathcal{R}(\mathbf r)\) in a patch centered on the average atomic-site position. The local pattern associated with this site displacement depends on the reciprocal-space components retained in the inverse transform. (a) \(\mathcal{R}(\mathbf r)\) obtained from the unmasked diffuse-scattering amplitude, featuring the most compact site-centered contrast dipole. (b) \(\mathcal{R}(\mathbf r)\) obtained by selecting an anisotropic subset of diffuse features exhibits a different shape for the positive and negative peaks, which are now asymmetric and significantly broader than in (a). (c) \(\mathcal{R}(\mathbf r)\) obtained by filtering the complementary reciprocal-space components exhibits a different spatial envelope. The MOSAIC $M$-decoder is designed to convert these patterns to their underlying atomic displacement vectors. Black arrows, originating at the average atomic-site position and scaled by \(2\times\) for visibility, indicate the displacement vectors recovered by the \(M\)-decoder from each local pattern.}
  \label{fig:Fig2}
\end{figure}

The central task here is to establish a quantitative mapping between specific reciprocal-space signatures and the underlying physical displacements of individual atoms. We refer to this displacement estimator as the Displacement Mode (Mode~2). The filtered real-space field $\mathcal{R}_{\mathcal{F}}(\mathbf{r})$ (Eq.~\eqref{eq:psi_feature}) does not directly yield displacement vectors $\mathbf{u}_s$. Physically, $\mathcal{R}_{\mathcal{F}}(\mathbf{r})$ represents a structural signal that is inherently delocalized.  This is because the reciprocal-space window $\mathcal{F}$ effectively acts as a band-pass filter, and the resulting real-space representation is a convolution of discrete atomic displacements with the window's point-spread function. Consequently, the information corresponding to a single atomic shift is distributed across multiple voxels and spatially overlapped with the signals of neighboring sites. An illustrative single-site response is shown in Fig.~\ref{fig:Fig2}.

To resolve this ambiguity, we treat the recovery of site-resolved displacements $\mathbf{u}_s$ as a localized inverse problem. We conceptualize $\mathcal{R}_{\mathcal{F}}(\mathbf{r})$ as a coherent superposition of site-centered responses. In an idealized limit, each site $s$ contributes a localized pattern, having a morphology determined by the window $\mathcal{F}$ and the average crystallographic structure, while the site displacement $\mathbf{u}_s$ modulates this pattern in a reproducible, predictable manner. The objective is to ``decode'' such complex, overlapping oscillations, mapping them to the displacement vectors from which they originated.

For small displacements, the displacement-sensitive part of the diffuse amplitude follows the first-order expansion
\begin{equation}
    \Delta A_{\mathrm{disp}}(\mathbf Q) \approx
    i\sum_j f_j(\mathbf Q)\,
    (\mathbf Q\cdot\mathbf u_j)
    \exp\bigl(i\mathbf Q\cdot\bar{\mathbf r}_j\bigr).
    \label{eq:first_order_disp}
\end{equation}
This expression motivates a linear inverse, but deriving a closed-form analytical decoder is often precluded by practical complexities, including non-uniform masking, finite sampling resolution, realistic form factors, and the intricate interference effects of multi-atomic bases. Therefore, we implemented a data-driven linear inverse. We refer to this operator as the $M$-decoder, which learns a fixed linear mapping of linearly preprocessed local patches (Supplementary Note~1.6) of  $\mathcal{R}_{\mathcal{F}}(\mathbf{r})$ onto the corresponding displacement vectors. By adopting this linear architecture, the method ensures that the decoded displacements remain strictly additive across disjoint regions of reciprocal space as detailed in Supplementary Note~1.

The displacement estimate is obtained as:
\begin{equation}
    \hat{\mathbf{u}}_s = M\,r_s,
    \label{eq:M_decoder_main}
\end{equation}
where $M$ is trained once using paired data $\{(r_s,\mathbf u_s^{\mathrm{true}})\}$ for the configuration being analyzed, where displacements are known, and then held fixed for inference (Supplementary Note~1). After training, $M$ is applied uniformly across all feature windows for that configuration; the dependence on $\mathcal F$ enters only through the input field $\mathcal R_{\mathcal F}$.

In the analysis presented here, $M$ was calibrated separately for each configuration using paired samples $\{(r_s,\mathbf u_s^{\mathrm{true}})\}$ derived from that same configuration, and was then held fixed when using different reciprocal-space masks. Thus, the additivity tests reported below probe the linearity of the masked-field decomposition and the stability of the fixed decoder, rather than cross-configuration generalization.

When $\mathcal R_{\mathcal F}(\mathbf r)$ differs systematically across distinct site types (e.g., corresponding to different Wyckoff positions in the average structure or chemically distinct coordination environments), a single global decoder can be overly restrictive. We therefore optionally use a small set of class-conditional decoders $\{M_{\alpha}\}_{\alpha=1}^K$ together with a fixed site-class assignment $\sigma(s)\in\{1,\dots,K\}$:
\begin{equation}
    \hat{\mathbf{u}}_s = M_{\sigma(s)}\,r_s.
    \label{eq:Ms_decoder_main}
\end{equation}
The mapping remains strictly linear, provided $\sigma(s)$ is specified independently of the masked $\mathcal R_{\mathcal F}(\mathbf r)$.

The linearity of the $M$-decoder holds for small displacements, where the resulting phase shifts can be expanded to first order. For larger displacements, the mapping becomes strongly non-linear, and the decoder may underestimate displacement magnitudes or miss aspects of the local response, although the spatial topology of the correlations is usually preserved.  In principle, the linear decoder can be generalized to a nonlinear map (e.g., a compact neural network that acts on $r_s$ or directly on the linearly preprocessed patch). Such models may improve accuracy in regimes with strongly environment-dependent responses, but they generally sacrifice strict additivity across disjoint reciprocal-space windows.  Therefore, here, we focus on the linear approximation.  

\subsection{Linearity and additivity over disjoint masks}
\label{subsec:linearity}

The mapping of the complex amplitude field $B(\mathbf{Q})$ to $\hat{\mathbf{u}}_s$ is linear at every stage: reciprocal-space windowing by $W_{\mathcal{F}}(\mathbf{Q})$, inverse transformation in Eq.~\eqref{eq:psi_feature}, site-centric patch extraction and spectral re-centering, inversion-odd extraction, fixed linear background removal, homogeneous linear feature construction $r_s=\mathcal{L}[\cdot]_s$, and linear decoding by $M$. Consequently, for two disjoint feature windows $\mathcal{F}_1$ and $\mathcal{F}_2$,
\begin{equation}
    \hat{\mathbf{u}}_s[B_1 + B_2] = \hat{\mathbf{u}}_s[B_1] + \hat{\mathbf{u}}_s[B_2],
\end{equation}
where $B_1(\mathbf{Q})$ and $B_2(\mathbf{Q})$ denote the corresponding masked amplitude fields. In the class-conditional case, Eq.~\eqref{eq:Ms_decoder_main}, the same statement holds, provided the class assignment $\sigma(s)$ is fixed independently of the feature window.

\section{Implementation }
\label{sec:implementation}

The mathematical framework outlined above establishes that site-resolved occupancy and displacement fields can be retrieved by filtering the diffuse scattering amplitude. Turning this formalism into a practical tool requires overcoming two main challenges: the computational difficulty of handling large 3D datasets and ensuring that the linear estimators are valid.

The implementation is released as the MOSAIC software package; availability details are provided in the Code availability statement.

\begin{figure}[!htbp]
  \centering
  \includegraphics[width=1.0\linewidth]{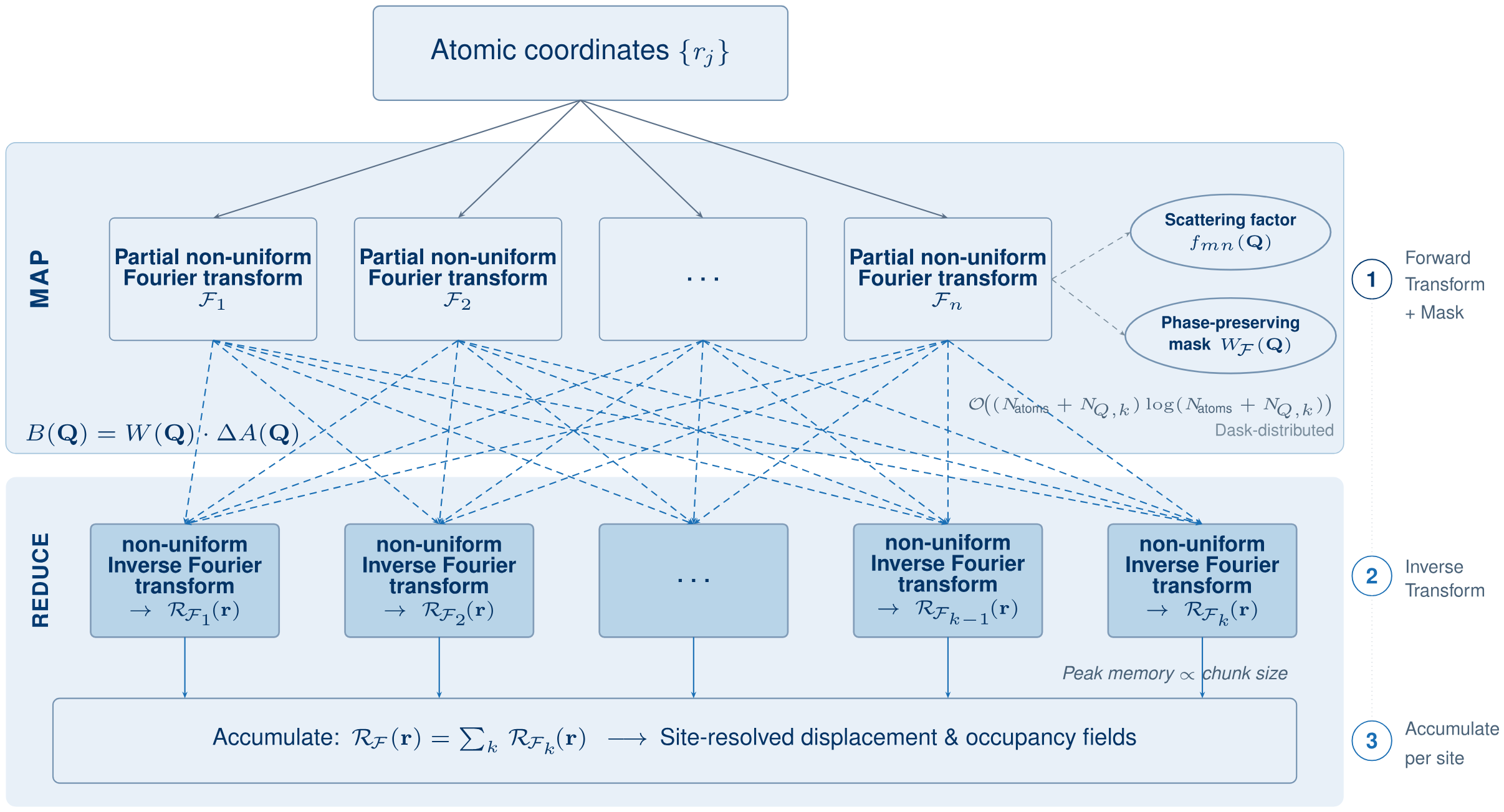}
  \caption{Schematic of the scalable Map--Reduce computational pipeline used in MOSAIC. Reciprocal space is partitioned into target subsets (called chunks) to manage the product scaling of direct transforms for large atomic configurations. For each chunk, a distributed type-3 NUFFT algorithm evaluates the scattering amplitude at arbitrary $\mathbf{Q}$-points.  Subsequently, a phase-preserving mask is applied, and a restricted inverse FT is computed in the neighborhood of atomic sites, eliminating the need for dense global 3D grids.}
  \label{fig:Fig3}
\end{figure}

\subsection{Scalable Architecture: The Map--Reduce Pipeline}
\label{subsec:map_reduce}

The principal bottleneck in implementing the MOSAIC framework is the size and structure of the FTs involved. Formally, both the forward transform to calculate the scattering amplitude, Eq.~\eqref{eq:Atot}, and the inverse transform to the structural-deviation field, Eq.~\eqref{eq:psi_feature}, are discrete FTs (DFTs) over finite sets of atoms and reciprocal-space points:
\begin{equation}
    A_{\mathrm{tot}}(\mathbf{Q}) = \sum_{j=1}^{N_{\text{atoms}}} f_j\,
    e^{i \mathbf{Q}\cdot\mathbf{r}_j}, \qquad
    R_F(\mathbf{r}) = \frac{1}{N_{\mathrm{tot}}}\sum_{\mathbf{Q} \in F}
    B(\mathbf{Q})\,e^{-i \mathbf{Q}\cdot\mathbf{r}}.
\end{equation}
Here, $F$ denotes the reciprocal-space feature mask (containing $N_Q=|F|$ points) and $N_{\mathrm{tot}}$ is a feature-independent normalization for the full transform volume as in Eq.~\eqref{eq:psi_feature}.
A direct evaluation of these sums scales as
\[
    O(N_{\text{atoms}} N_Q).
\]
That is, doubling either the number of atoms or the number of $\mathbf{Q}$-points in the masked region doubles the cost of every individual sum. For supercells with $10^6$ atoms and mask sets that often require $10^7$--$10^8$ $\mathbf{Q}$-points, this product scaling makes computations prohibitively expensive.

In principle, a standard FFT reduces the cost to $O(N \log N)$, but only when both the source and the target are on uniform grids. In MOSAIC, neither side of the transform satisfies this condition.  First, atomic coordinates $\{\mathbf{r}_j\}$ are arbitrary positions represented as floating-point values rather than nodes of a regular lattice.  Second, reciprocal-space points, at which $A(\mathbf{Q})$ and $B(\mathbf{Q})$ have to be calculated, are defined by user-constructed masks $F$ that follow the shapes of specific diffuse features rather than a dense Cartesian cube.

One could place atoms onto a very fine, regular real-space grid and compute a global 3D FFT up to $Q_{\max} \approx 50$~\AA$^{-1}$, but this would require  $\gtrsim 10^{9}$ voxels and hence hundreds of gigabytes to several terabytes of memory for a complex-valued $A(\mathbf{Q})$, despite only a small subset of that reciprocal-space volume being used by the masks.

To avoid both the product scaling of the direct DFT and the memory requirements of a global fine grid, MOSAIC treats the problem as a nonuniform FT in both real and reciprocal spaces, and evaluates it using NUFFT algorithms of type~3 (illustrated in Fig.~\ref{fig:Fig3}) ~\cite{greengard2004accelerating,barnett2019parallel}. These algorithms retain the near-FFT computational complexity, while affording the use of arbitrary source positions $\mathbf{r}_j$ and arbitrary target points $\mathbf{Q}$ selected by the masks.

We implemented this method within a high-throughput, out-of-core architecture based on the Map--Reduce paradigm, exploiting the linearity of both the forward and inverse transforms. The target reciprocal-space selection $F$ is partitioned into a set
of disjoint subregions $\{F_k\}$ (hereafter, referred to as ``chunks''). The real space field $\mathcal{R}_{\mathcal{F}}(\mathbf{r})$  is described as
\begin{equation}
    R_F(\mathbf{r}) = \sum_k R_{F_k}(\mathbf{r}), \qquad
    R_{F_k}(\mathbf{r}) = \frac{1}{N_{\mathrm{tot}}} \sum_{\mathbf{Q} \in F_k}
    B(\mathbf{Q}) e^{-i \mathbf{Q}\cdot\mathbf{r}},
\end{equation}
where $F_k$ contains $N_{Q,k}$ reciprocal-space points and $B(\mathbf{Q})$ is the diffuse (or total) amplitude selected by the mask.

The computation proceeds as described in Supplementary Note~1.

Each stage of the process --  computing $A_{\mathrm{tot}}(\mathbf{Q})$ via NUFFT, applying reciprocal-space masks, mapping back to atomic sites with the inverse NUFFT -- is strictly linear in the input amplitude. Therefore, the final displacement and occupancy fields do not depend on how $F$ is partitioned into chunks. The streaming Map–Reduce architecture decouples peak memory usage from the overall problem size: at any time, only the current $\mathbf{Q}$-chunk and a small buffer of atomic-site-centered real-space values are stored in memory. This makes the analysis of large configurations and long molecular-dynamics trajectories practical while preserving the linearity and phase sensitivity required by the MOSAIC estimators.

\section{Validation}

\begin{figure}[!htbp]
  \centering
  \includegraphics[width=0.72\linewidth]{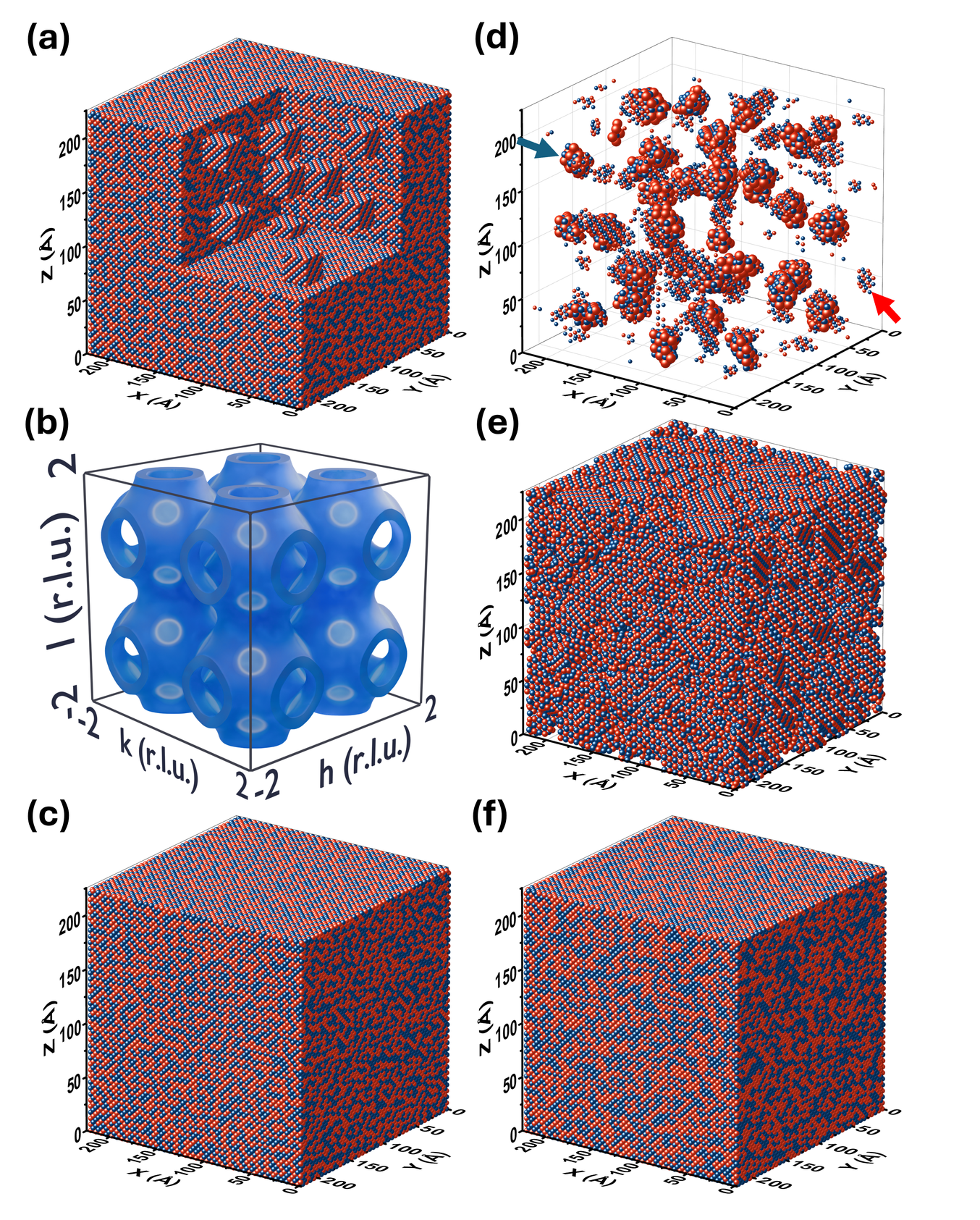}
\caption{Application of MOSAIC to a simulated $\gamma$-LiFeO$_2$ structure with the coexisting Li/Fe short-range order and nanoscale $L1_1$ ordering. \textbf{(a)} Structural model containing $40\times40\times40$ unit cells with cubic nanoregions exhibiting the $L1_1$-type Li/Fe superstructure, embedded in a matrix featuring a short-range order of these species. The $L1_1$ superstructure in the nanoregions, shown without the matrix in the cutout, exhibits 30\% antisite disorder.   Li and Fe are represented in blue and red, respectively.  \textbf{(b)} Diffuse-intensity surface arising from the short-range order in the matrix. The $L1_1$ ordering in the nanoregions contributes additional intensity at the $\tfrac{1}{2}111$-type points, marked by light-shaded disks. \textbf{(c)} Reconstruction of atomic site occupancies in the matrix via the inverse Fourier transform of the scattering amplitude associated with the diffuse surface in \textbf{(b)}, isolated using a custom mask while excluding spherical regions around  $\tfrac{1}{2}111$.   The reconstruction reveals the Li/Fe configurations that form the short-range order, excluding those associated with the $L1_1$ superstructure. \textbf{(d)} Reconstruction similar to \textbf{(c)}, but obtained using spherical masks centered at the $\tfrac{1}{2}111$-type positions.  Only atoms with the transform intensities above a certain threshold are displayed.   Larger clusters represent the  $L1_1$-ordered nanoregions, whereas the smaller clusters correspond to the  $L1_1$-like Li/Fe motifs in the matrix that also contribute to the $\tfrac{1}{2}111$-type reflections. \textbf{(e)} The same reconstruction as in \textbf{(d)},  but with the transform intensity adjusted to highlight the Li/Fe motifs in the matrix associated specifically with the diffuse intensity at $\tfrac{1}{2}111$. \textbf{(f)} Reconstruction from the diffuse signal outside the surface shown in  \textbf{(b)}. Overall, this figure illustrates MOSAIC's ability to separate atomic configurations that generate diffuse scattering at distinct locations in reciprocal space.}
  \label{fig:Fig4}
\end{figure}

\subsection{Chemical Ordering}
\label{subsec:LiFeLiFeO2}

Solid solutions may exhibit competing types of short-range chemical order with distinct correlation lengths. Here, we examine a common case of short-range order in rocksalt-type structures $A(B_1)_x(B_2)_{1-x}$, where the B-sites are occupied by a mixture of B1 and B2 species, arranged so that all local octahedral $[AB_6]$ coordination environments have the same stoichiometric B1/B2 ratio. Examples of systems with such SRO include transition-metal carbides, such as $VC_{1-x}$ (here, x is the vacancy content), and oxides, such as LiFeO$_2$. We considered a situation in which this SRO coexists with the nanoscale ordering of B1 and B2, characterized by the ordering vector $\mathbf{k}=\tfrac{1}{2}\langle111\rangle^{*}$, and used a simulated structure to test MOSAIC’s ability to disentangle these two effects.  

\textbf{Model construction and reciprocal space signatures.}  Specifically, we considered a hypothetical model of  $\gamma$-LiFeO$_2$, which is known to exhibit complex diffuse scattering manifolds driven by local charge-neutrality constraints and Li/Fe ordering~\cite{ji2019hidden,li2023atomic,sauvage1974prediction}. We generated an atomic configuration consisting of $40 \times 40 \times 40$ unit cells.  Li and Fe were distributed using an MC algorithm that enforced a local coordination rule requiring every oxygen to be coordinated by exactly 3 Li and 3 Fe cations. This constraint creates SRO, manifested in a characteristic, continuous diffuse intensity surface defined by the equation $\cos(\pi h) + \cos(\pi k) + \cos(\pi l) \approx 0$~\cite{sauvage1974prediction,ji2019hidden}. The resulting configuration contains no LRO. We then introduced cubic nanoregions (edge length $\approx 2$~nm) exhibiting the L$1_1$ superstructure  ($\alpha$-LiFeO$_2$ type), with Li and Fe segregated into alternating (111) planes, albeit with a significant degree of antisite disorder.  This nanoscale ordering, which results in trigonal symmetry, is based on one of the two inequivalent Li/Fe configurations around oxygen (e.g., one triangular face Li-Li-Li and the opposite face Fe-Fe-Fe).   The scattering amplitude calculated for this composite model still exhibits the same diffuse-scattering surface, reflecting the SRO, but with some condensation of diffuse intensity at $\tfrac{1}{2}111$-type ($L$-point) locations, associated with the L$1_1$ superlattice within the nanodomains. Because the average composition and the local 3:3 coordination are identical in the matrix and the clusters, it is not straightforward to distinguish between these two components through a direct real-space analysis of the cation distributions.  As we show below, the filtering approach in MOSAIC provides an effective solution to this problem.

\textbf{MOSAIC reconstruction.} We analyzed the ordering of Li and Fe using the \textbf{Chemical Mode} filter and two complementary reciprocal-space mask sets. First, a spherical mask was applied to the $\tfrac{1}{2}111$-type reflections (the ``Nanodomain Filter''). The resulting MOSAIC real-space reconstruction highlights atomic arrangements with the L$1_1$-type ordering, while suppressing other configurations.  Displaying atoms that acquire intensity in the inverse FT above a certain threshold reveals dense nanoscale clusters with the well-defined L$1_1$ superstructure, along with smaller fragments also featuring the L$1_1$ motif.  The cubic shape of the clusters is not reproduced due to significant Li/Fe disorder within these regions, which blurs the cluster/matrix interface.  The smaller fragments reflect parts of the matrix with Li and Fe arrangements containing alternating Li-Li-Li and Fe-Fe-Fe \(\{111\}\) planes, which also contribute diffuse intensity at the\textit{ L} points.  As in the original configuration, the nanoclusters contain a single variant of the superstructure.  In contrast, the matrix incorporates local atomic configurations corresponding to all four $\langle111\rangle$ variants, and these are reproduced after filtering. Second, a complementary mask was applied to the cosine surface, excluding the superlattice reflections. 
The corresponding reconstruction (Fig.~\ref{fig:Fig4}) reveals the matrix SRO configurations, whereas motifs associated with the L$1_1$ superstructure, including atoms within the L$1_1$-ordered nanoclusters, exhibit near-zero intensity in the inverse FT. Thus, chemical SRO contributing diffuse scattering at specific locations in reciprocal space can be readily identified and separated using MOSAIC. For similar SROs with different correlation lengths, the corresponding contributions remain distinguishable, but their interpretation requires additional analysis.

\subsection{Atomic Displacements}
\label{subsec:ReO3}

\begin{figure}[!htbp]
  \centering
  \includegraphics[width=1.0\linewidth]{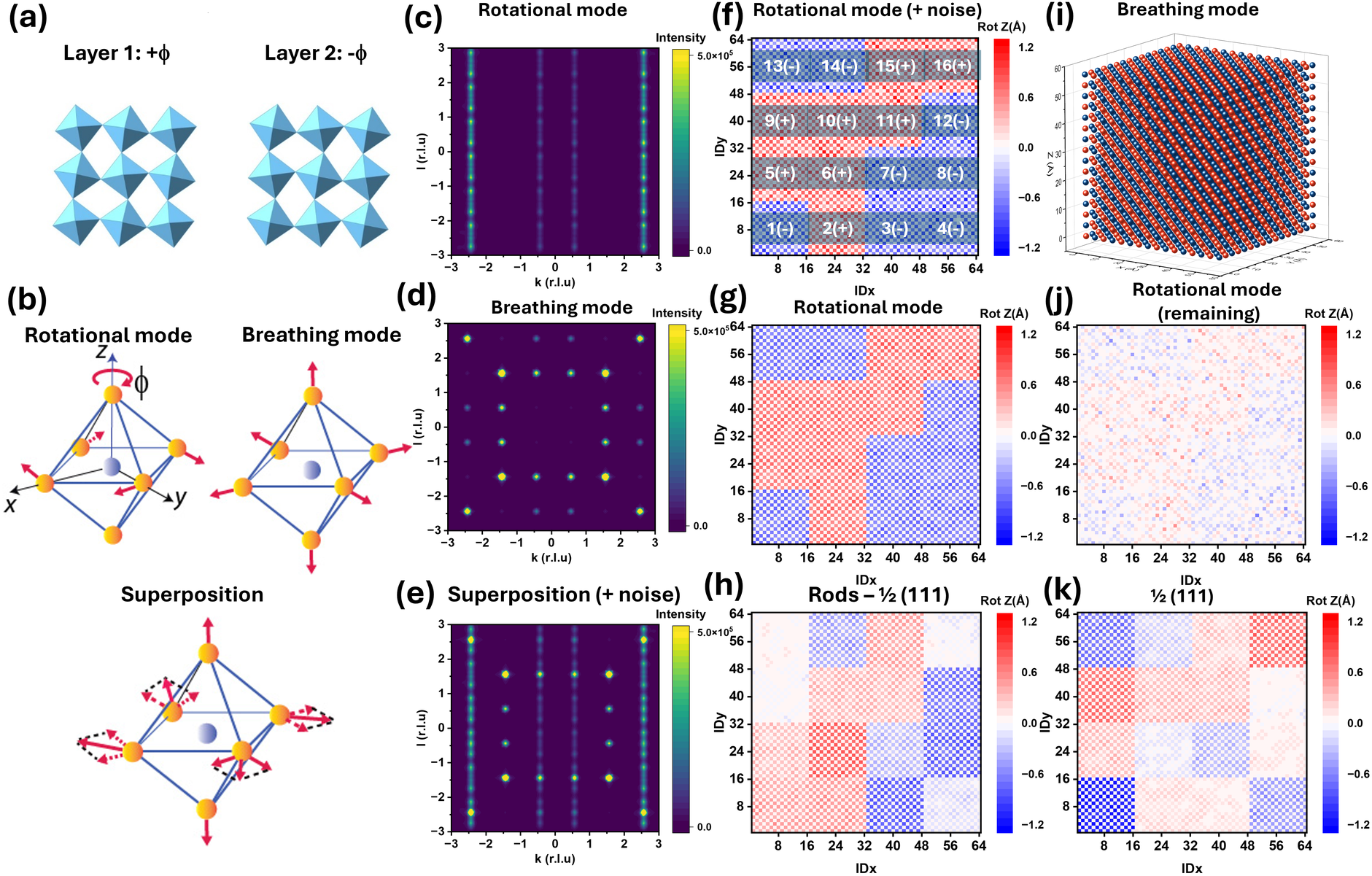}
\caption{Disentangling overlapping diffuse-scattering signatures of rotational and breathing distortions in a perovskite-like octahedral framework. \textbf{(a)} Layers of vertex-sharing octahedra featuring out-of-phase $+\phi$ and $-\phi$ rotations. \textbf{(b)} Schematic representation of the rotational and breathing distortion modes,  and their superposition. \textbf{(c--e)} Simulated diffuse-scattering intensity ($k$--$l$ sections, logarithmic scale) for the following models: \textbf{(c)} randomly stacked layers 1 and 2; \textbf{(d)} long-range ordering of the breathing distortion forming the rocksalt-type arrangement of expanded and contracted octahedra; and \textbf{(e)} both distortions combined.   The octahedral rotations give rise to diffuse rods along $[\tfrac{1}{2},\tfrac{1}{2},l]$, parallel to the rotation axis, whereas the ordered breathing distortion produces diffuse maxima at the $\tfrac{1}{2}111$-type positions along these rods. \textbf{(f)}  Map of octahedral rotations, $R_z$, in the model, showing 16 octahedral layers, arranged in a $4 \times 4$ array, with layer indices indicated.  Symbols represent every other octahedron in each layer, and the color (blue or red) denotes the sign of rotation as in (a). The rotations of the remaining octahedra are determined by connectivity. \textbf{(g)} Reconstruction of the rotational field via inverse Fourier filtering using the entire diffuse rods. \textbf{(h)} Same as in \textbf{(g)}, but excluding spherical regions around the $\tfrac{1}{2}111$ reflections. \textbf{(i)} Reconstruction of the breathing distortion using spherical masks centered on $\tfrac{1}{2}111$ reflections; colors indicate expanded (red) and contracted (blue) octahedra.  \textbf{(j)} Residual rotational field after subtraction of the full-rod-based reconstruction. \textbf{(k)} Reconstruction analogous to \textbf{(g)} but using the complementary filtering described in \textbf{(i)}.  The sum of \textbf{(h)} and \textbf{(k)} reproduces \textbf{(g)}, confirming the additivity of filtered displacements when complementary masks are used.}
  \label{fig:Fig5}
\end{figure}

A second benchmark targets coexisting structural distortions that generate overlapping diffuse-scattering features. As a model system, we use a simulated ReO$_3$-type perovskite-like structure with a framework of corner-sharing octahedra.  The purpose of this example is to demonstrate that MOSAIC can separate components of atomic displacements engaged in distinct distortion modes (Fig.~\ref{fig:Fig5}).

\textbf{Model construction.}
We considered a configuration containing $16\times16\times16$ cubic perovskite unit cells with a lattice parameter $a = 4$~\AA.  For specificity, we adopted a [BO$_6$] octahedral framework, leaving the cubeoctahedral sites vacant. Two types of distortions were introduced.  One is a rotational mode around one of the cubic axes (z-axis in Fig.~\ref{fig:Fig5}).  The rotation amplitude was set to $\phi = 10^\circ$. Because of the corner connectivity, the octahedra in each $z$ layer form a cogwheel pattern of $\pm\phi$ rotations. For successive octahedra along the z-axis, the rotations are unconstrained.  We considered a case where the sign of the rotations along this axis varies randomly, so that the structure consists of a random sequence of Type 1 and Type 2 layers shown in Fig.~\ref{fig:Fig5}.  (see also Ref.~\cite{Eremenko2017}.)   This type of disordered rotation produces rods of diffuse intensity that extend parallel to the [001] direction through $\tfrac{1}{2}hk$ ($h,k$ odd) positions (Fig.~\ref{fig:Fig5}).  In our simulations, the rods appear spotty because of the limited statistics afforded by the 16 octahedral layers in the configuration. 

The second mode we introduced is a breathing-type distortion, that results in a uniform expansion (Fig.~\ref{fig:Fig5}) or contraction of the octahedra.  We set the magnitude of the displacements of the ligand atoms in this mode to $\Delta r = 0.10$~\AA. We varied the sign of this distortion to introduce a long-range rocksalt-type ordering of the expanded and contracted octahedra, with this ordering giving rise to sharp $\tfrac{1}{2}111$-type reflections, superimposed on the diffuse rods from the tilting disorder.

All atoms were subjected to additional random positional disorder with isotropic Gaussian displacements having mean-squared magnitudes of $U_{\mathrm{B}} = U_{\mathrm{O}} = 0.005$~\AA$^{2}$.

\textbf{Reciprocal-space masks.}
To analyze the origins of the diffuse scattering for this configuration, we defined two complementary sets of masks. The first set was cylinders that enclose the $[\tfrac{1}{2},\tfrac{1}{2},l]$-type rods, but without spherical regions around $\tfrac{1}{2}111$:
    \[
    \begin{aligned}
        \bigl(\mathrm{Mod}(h,1)-\tfrac{1}{2}\bigr)^{2}
      + \bigl(\mathrm{Mod}(k,1)-\tfrac{1}{2}\bigr)^{2}
      &\le r_1^{2}, \\
        \bigl(\mathrm{Mod}(h,1)-\tfrac{1}{2}\bigr)^{2}
      + \bigl(\mathrm{Mod}(k,1)-\tfrac{1}{2}\bigr)^{2}
      + \bigl(\mathrm{Mod}(l,1)-\tfrac{1}{2}\bigr)^{2}
      &\ge r_2^{2},
    \end{aligned}
    \]
with $r_1 = 0.1876$ and $r_2 = 0.2501$ in reciprocal-lattice units.  The second set was spheres centered on the $\tfrac{1}{2}111$-type points:
    \[
        \bigl(\mathrm{Mod}(h,1)-\tfrac{1}{2}\bigr)^{2}
      + \bigl(\mathrm{Mod}(k,1)-\tfrac{1}{2}\bigr)^{2}
      + \bigl(\mathrm{Mod}(l,1)-\tfrac{1}{2}\bigr)^{2}
        < r_2^{2}.
    \]
The resulting filtered amplitude $\Delta A(\mathbf{Q})$ was decoded using the Displacement Mode estimator to obtain the site-resolved displacement fields $\mathbf{u}_{\mathrm{rod}}(\mathbf{r})$ and $\mathbf{u}_{\mathrm{sphere}}(\mathbf{r})$.

\textbf{MOSAIC reconstruction and error analysis.}
Unlike chemical ordering, analyzing filtered displacements in terms of distortion modes requires converting decoded site displacements into mode amplitudes, following the procedure used in Ref.~\cite{Eremenko2019}. Here, we use the same approach to extract octahedral rotations and breathing distortions. Applying the cylindrical and spherical masks together (i.e., filtering the full diffuse scattering) recovers the octahedral rotations, including their patterns and magnitudes, while suppressing statistical noise introduced by the added random displacements (Fig.~\ref{fig:Fig5}(g)). In contrast, applying only the spherical masks recovers the breathing mode, reproducing its ordered arrangement of expanded and contracted octahedra and the magnitude of the associated displacements (Fig.~\ref{fig:Fig5}(i)). Filtering using the cylindrical masks while excluding the spheres around $\tfrac{1}{2}111$ produced a complementary rotation map, also shown in Fig.~\ref{fig:Fig5}(h). Summing these two maps reconstructs the original rotation field, confirming the linear additivity of the filtered displacements obtained with complementary masks. The residual 
\[
    \Delta\mathbf{u}(\mathbf{r}) =
    \bigl[\mathbf{u}_{\mathrm{rod}}(\mathbf{r})
         + \mathbf{u}_{\mathrm{sphere}}(\mathbf{r})\bigr]
    - \mathbf{u}_{\mathrm{full}}(\mathbf{r})
\]
is negligible compared with the characteristic displacement scale ($\approx 0.1$~\AA). Quantitative validation of this 3D reconstruction is shown in Supplementary Fig.~S5, where the additivity residual $|\mathbf{u}_{\mathrm{sphere}}+\mathbf{u}_{\mathrm{rod}}+\mathbf{u}_{\mathrm{rest}}-\mathbf{u}_{\mathrm{all}}|$ and the reconstruction residual $|\mathbf{u}_{\mathrm{all}}-\mathbf{u}_{\mathrm{true}}|$ are presented. This demonstrates that MOSAIC can disentangle coexisting lattice distortions involving the same atoms, provided they are distinct in character.

\section{Results: Applications to Experimental Models}
\label{sec:results}

In this section, MOSAIC is applied to analyze atomistic models of  relaxor ferroelectrics, \allowbreak{} PbMg$_{1/3}$Nb$_{2/3}$O$_3$ (PMN) and 0.7PMN--0.3PbTiO$_3$ (PMN-PT).  These materials crystallize in a perovskite structure and exhibit coupled chemical and displacement correlations spanning several length scales that give rise to rich diffuse scattering \cite{Bokov2006,hlinka2012we,rojac2023piezoelectric}.  Recently, comprehensive atomistic models of these structures were obtained from RMC refinements which simultaneously fit X-ray and neutron total scattering, EXAFS, and 3D diffuse scattering data (\cite{Eremenko2019,eremenko2025emergent}). Here, we use these models to demonstrate the capabilities of MOSAIC. 

\subsection{Chemical short-range order }
\label{subsubsec:pmn_chemical}

 PMN exhibits chemical SRO for Mg and Nb, which share the octahedral B sites. This ordering adopts the rocksalt-type arrangement that can be described as $\mathrm{Pb}(B'_{1/2}B''_{1/2})\mathrm{O}_3$, in which $B'$ (Mg/Nb) and $B''$ (Nb) sites occupy alternating \(\{111\}\) planes.   The ordering is confined to the nanoscale and gives rise to diffuse intensity maxima at half-integer (h,k)-type positions, where $h,k,l$ are all odd indices.   The degree of ordering varies continuously throughout the material, making direct visualization of these variations in large-scale atomistic structural models challenging.

We first examine two-dimensional views of these 3D configurations, starting with {110} projections.  In this orientation, \emph{B}$'$ and \emph{B}$''$ sites are separated into distinct atomic columns, which aids visualization.  Still, a substantial degree of disorder obscures this picture.  The situation is particularly challenging in PMN-PT, where the ordering is weaker and the B sites are occupied by three different species, as shown in Supplementary Fig.~S1.

The diffuse-intensity patterns for the {110} projection exhibit peaks at $\tfrac{1}{2}(hk)$ positions.\cite{cowley2011relaxing}  Applying MOSAIC's Chemical Mode to the total, unmasked diffuse amplitude yields an inverse FT that reproduces the original Mg/Nb arrangement.  If the filtering is restricted to the $\tfrac{1}{2}hk$ features, the resulting atomic-column intensities reflect deviations of local scattering power from the average.  For X-rays, positive and negative deviations correspond to Mg-rich (\emph{B}$'$ sites) and Nb-rich (i.e., \emph{B}$''$ sites) columns, respectively.  The magnitude of these deviations scales with the departure from randomness, reaching maxima for \emph{B}$'$=Mg and \emph{B}$''$=Nb.

\begin{figure}[!htbp]
  \centering
  \includegraphics[width=0.7\linewidth]{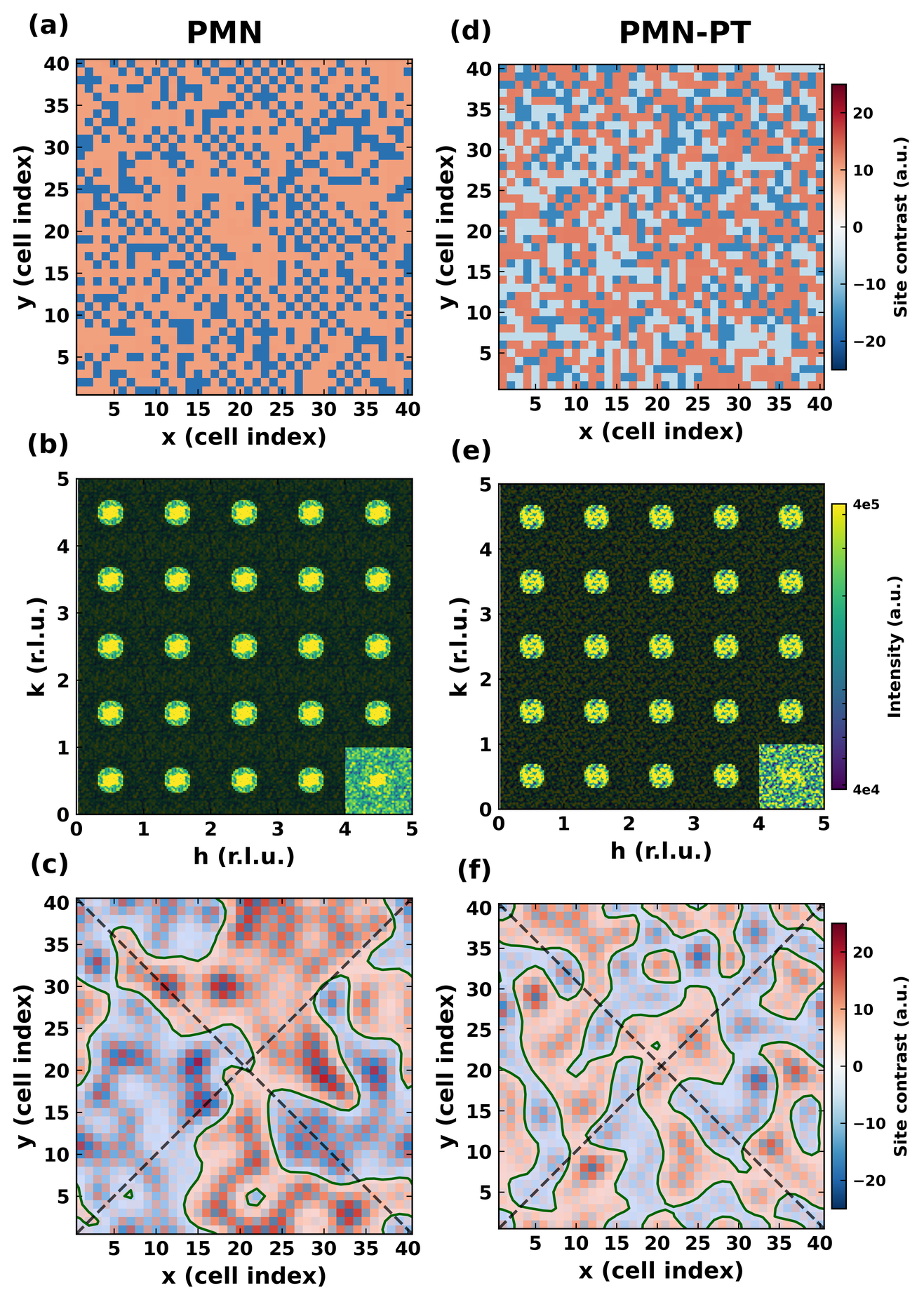}
  \caption{Chemical short-range order in PMN and PMN-PT is visualized via the{001} planar slices. \textbf{(a),\,(d)} Single {001} layer in the structural models of PMN and PMN-PT, respectively, obtained using RMC refinements against experimental data. Color indicates deviations of the site occupancy from its average value. Red and blue colors correspond to the Nb-rich and Nb-poor sites, respectively.
  \textbf{(b),\,(e)}~Diffuse-scattering patterns calculated from the slices in (a) and (d), respectively. The intensity is displayed on a logarithmic scale. The patterns exhibit peaks at half-integer positions, indicating chemical ordering. For PMN, extended patches with chessboard patterns are readily observed in (a), giving rise to relatively strong and sharp peaks in (b). For PMN-PT, the occupancy distribution in (d) appears mostly random despite the diffuse yet distinct half-integer peaks in (e) pointing to the existence of chemically ordered regions.
  \textbf{(c),\,(f)}~MOSAIC-filtered maps obtained by applying circular masks to the half-integer peaks as shown in (b,e). The sign of the reconstructed field distinguishes 
  $B'$-type from $B''$-type sublattice assignment, while its 
  magnitude tracks the local ordering strength. Green contour lines 
  mark the anti-phase boundaries. Dashed diagonal lines are guides to the 
  eye highlighting the preferential $\langle 110 \rangle$ alignment of the 
  domain boundaries. The PMN configuration exhibits relatively large, 
  well-defined ordered domains, whereas PMN-PT shows smaller, more 
  fragmented regions, consistent with reduced chemical ordering upon the Ti substitution.}
  \label{fig:Fig6}
\end{figure}

For isolated {001} slices (Fig.~\ref{fig:Fig6}, top row) of the as-refined PMN configuration, where the ordering is relatively well-developed, chessboard-like patches are visible even without filtering, yielding pronounced superlattice reflections.  In contrast, for PMN-PT, the distribution appears highly stochastic, yet still yields broad, weak superlattice spots distinguishable from the background.  Applying MOSAIC filtering to these peaks (Fig.~\ref{fig:Fig6}, middle row) generates clear maps of the state of ordering, revealing a labyrinthine structure of strongly and weakly ordered regions. Some of the latter form wide anti-phase boundaries separating two translation variants of the superlattice.  

Applying the Chemical Mode  to the 3D PMN and PMN-PT configurations,  using spherical masks centered on the $\tfrac{1}{2}111$-type diffuse peaks, reveals a continuously varying degree of rocksalt-type ordering rather than distinct ordered regions within a disordered matrix.  Regions of stronger order appear as fluctuations in the local order parameter separated by wide regions, often anti-phase boundaries, where the ordering varies gradually, as supported by prior TEM observations. \cite{pasciak2012polar,krogstad2018relation,cabral2018gradient,eremenko2025emergent,reaney1994b}

These results highlight the main practical advantage of MOSAIC filtering: site- or column-resolved classification into distinct chemical environments is obtained directly from the phase and sign of the filtered field, overcoming spatial and chemical blurring without the need for explicit clustering algorithms or forward modeling.

\subsection{Polar textures}

\begin{figure}[!htbp]
  \centering
  \includegraphics[width=\linewidth]{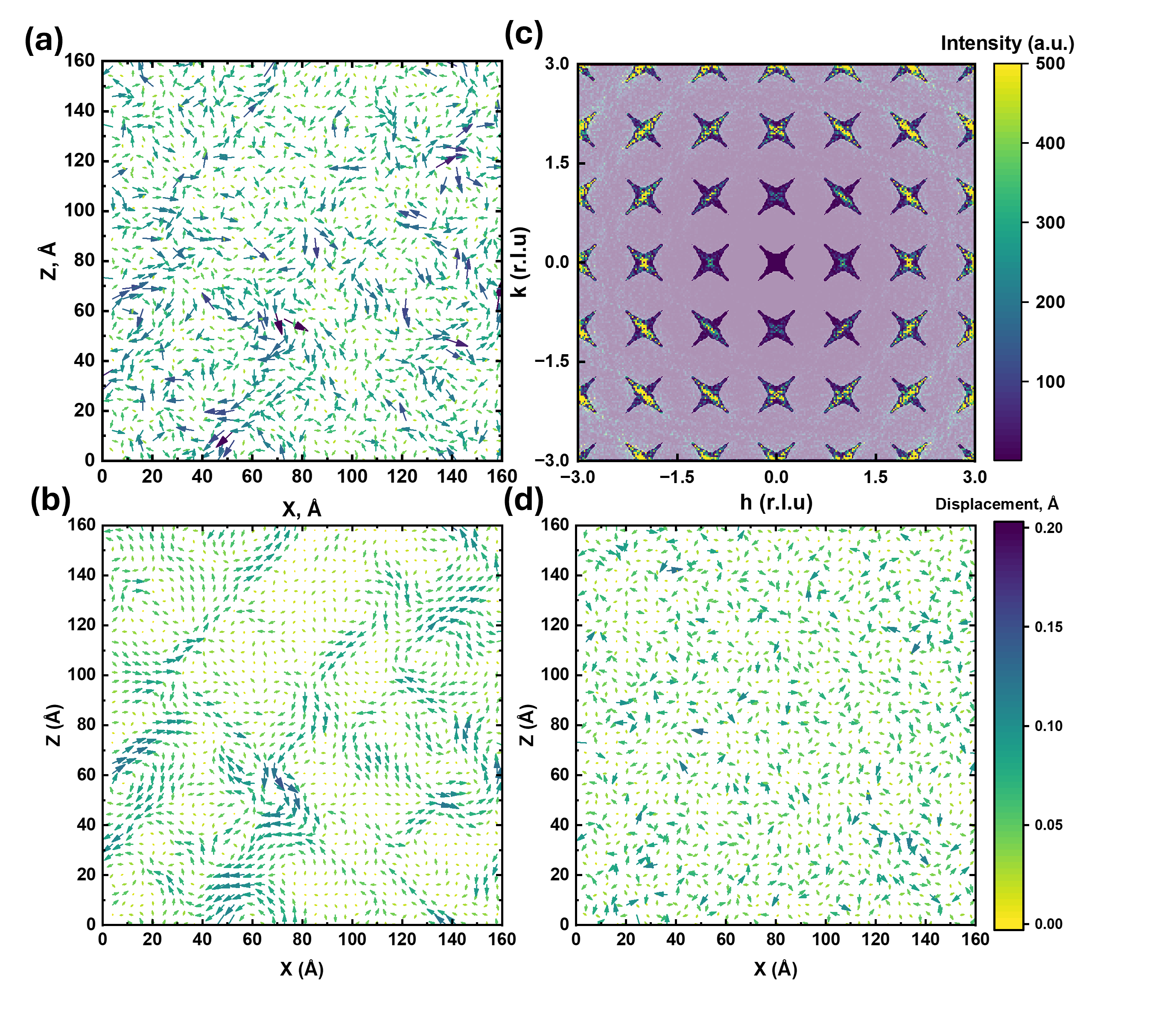}
  \caption{Analysis of a  2D projection of a PMN supercell.
  (a) Pb-column displacement vectors obtained by projecting the configuration along the cubic axis and measuring each Pb-column shift relative to the ideal lattice (vectors scaled $\times 50$ for visibility); the unfiltered map is dominated by uncorrelated noise.
  (b) Corresponding calculated 2D diffuse scattering intensity in the $(h,k)$ plane (arb.~units), showing butterfly-shaped diffuse wings around the Bragg positions.
  (c) Displacement map reconstructed by applying the phase-preserving filtering to the Fourier representation of the projected Pb-column displacement field, retaining only the butterfly-shaped diffuse wings, and inverting the masked signal with the same displacement estimator.
  (d) Complementary reconstruction from the remaining Fourier components (butterfly wings removed), showing that non-butterfly contributions primarily contribute an uncorrelated noise background.}
  \label{fig:Fig7}
\end{figure}

Here, we analyze the same PMN and PMN--PT configurations using the Displacement Mode (Fig.~\ref{fig:Fig7}).  These configurations reproduce the characteristic, anisotropic diffuse scattering in the vicinity of Bragg peaks. 

In three dimensions, the unfiltered Pb-displacement field extracted directly
from the as-refined RMC configurations appears noisy, obscuring visual detection of correlations.  Applying the Displacement Mode to the diffuse amplitude isolated using spherical masks centered on the Bragg peaks yields a reconstructed Pb field revealing that Pb displacements are preferentially aligned within extended volumes along a small subset of crystallographic directions. The planar projections exhibit swirling patterns of polarization, also containing vortices, as shown in
Ref.~\cite{eremenko2025emergent}.  

To mimic the displacement fields typically extracted from STEM images, we generated a (001) projection from our 3D configuration and identified displacements of Pb columns from the average positions (Fig.~\ref{fig:Fig7}).  The corresponding 2D diffuse scattering pattern exhibits characteristic butterfly-shaped features around the Bragg peaks, indicating the presence of extended correlations.  Before filtering, however, the displacement field is dominated by uncorrelated noise that obscures correlation patterns.  Applying the Displacement Mode filter, with diffuse scattering isolated by circular masks centered on the Bragg peaks and enclosing the diffuse features, removes the random components and yields a clear visualization of the correlations underlying the diffuse signal.  
For this projected-column example (Fig.~\ref{fig:Fig7}), the phase-bearing input is the Fourier representation of the measured displacement field rather than a full kinematic scattering amplitude.

The combined examples of chemical and polar correlations in PMN illustrate that MOSAIC provides a reproducible, model-conditioned attribution of specific scattering features
to their corresponding site-resolved fields, applicable to both three-dimensional atomic configurations and two-dimensional structural projections. For chemical short-range order, the filtered field serves as a proxy for a local rocksalt-like order parameter, directly identifying \emph{B$'$}- and \emph{B$''$}-like environments, even in the presence of considerable disorder. For displacement correlations, the same filtering approach reveals nanoscale polar textures despite the significant noise in the original configuration.

\section{Summary}
\label{sec:discussion}

We present a quantitative method and software framework, called MOSAIC, for attributing specific diffuse-scattering features to atomic site-occupancy and the displacement fields from a known atomic configuration.   The method uses phase-preserving Fourier filtering of the scattering amplitude, with the inverse Fourier transform evaluated at or near atomic positions in the configuration of interest to determine site-resolved deviations from the average structure. 

A central component of MOSAIC is a linear decoder that maps the inverse transform to site occupancy and displacements, enabling separation of distinct structural distortions and chemical ordering that may produce overlapping scattering features. The software leverages the Map--Reduce architecture and non-uniform fast Fourier transforms to handle large, million-atom configurations and extensive regions of reciprocal space while avoiding the memory limitations of traditional grid-based approaches.  This method is applicable to both three-dimensional structural models derived from experimental data or computer simulations, and two-dimensional structural projections, such as those obtained from atomic-resolution transmission electron microscopy images.  

The computational scalability of MOSAIC opens the possibility of applying it to other types of analysis. One example is 4D-STEM, in which a full convergent-beam electron diffraction pattern is recorded at each probe position during a scan. In this context, MOSAIC could be adapted to apply phase-preserving filters and linear decoders within an appropriate image-formation or structural-reconstruction model to map local order parameters from experimental data.  Another application is the analysis of diffuse scattering from individual snapshots in MD trajectories, enabling efficient tracking of displacement fields associated with specific vibrational modes. This may help link diffuse-scattering features to underlying material properties.

\section{Code availability}
The MOSAIC source code is available at
\url{https://github.com/MaximEremenko/MOSAIC.git}.
The repository contains the implementation used for reciprocal-space masking,
restricted inverse Fourier transforms, and site-resolved chemical and displacement
reconstructions described in this work.
\section{Acknowledgments}
\label{sec:Acknowledgments}
This work was performed under the following financial assistance award 70NANB24H134
from U.S. Department of Commerce, National Institute of Standards and Technology. This work was performed under subcontract CW44332 from UT-Battelle, LLC, c/o Oak Ridge National Laboratory, a subcontract under Contract DE-AC05-00OR22725 from the U.S. Department of Energy.

\section*{Correspondence and requests for materials}
Correspondence and requests for materials should be addressed to M.E. (eremenkom@ornl.gov).

\bibliographystyle{unsrt}
\bibliography{bibl}
\end{document}